\documentclass[12pt]{article}
\usepackage{geometry}
\usepackage{graphicx}
\usepackage{amsmath}
\usepackage{amsfonts}
\usepackage{amssymb}
\usepackage{cite}
\usepackage{epstopdf}
\usepackage{comment}
\usepackage{cancel}
\usepackage{mathrsfs}
\usepackage{dsfont}

\usepackage{cleveref}
\usepackage{cite}
\crefname{section}{§}{§§}
\Crefname{section}{§}{§§}
\newcommand{\mfrac}[2]{{#1/#2}}
\numberwithin{equation}{section}
\usepackage[titletoc,title]{appendix}

\usepackage{booktabs} 

\newcommand{\bea}{\begin{eqnarray}}

\newcommand{\eea}{\end{eqnarray}}

\newcommand{\be}{\begin{equation}}
\newcommand{\ee}{\end{equation}}
\newcommand{\ba}{\begin{align}}
\newcommand{\ea}{\end{align}}

   \makeatletter
  \let\over=\@@over \let\overwithdelims=\@@overwithdelims
  \let\atop=\@@atop \let\atopwithdelims=\@@atopwithdelims
  \let\above=\@@above \let\abovewithdelims=\@@abovewithdelims
\renewcommand\section{\@startsection {section}{1}{\z@}%
                                   {-3.5ex \@plus -1ex \@minus -.2ex}
                                   {2.3ex \@plus.2ex}%
                                   {\normalfont\large\bfseries}}

\renewcommand\subsection{\@startsection{subsection}{2}{\z@}%
                                     {-3.25ex\@plus -1ex \@minus -.2ex}%
                                     {1.5ex \@plus .2ex}%
                                     {\normalfont\bfseries}}

\newcommand{\ophi}[0]{\overline{\phi}}
\newcommand{\oPhi}[0]{\overline{\Phi}}
\newcommand{\oF}[0]{\overline{F}}
\newcommand{\of}[0]{\overline{f}}
\newcommand{\oh}[0]{\overline{h}}
\newcommand{\oeta}[0]{\overline{\eta}}
\newcommand{\oell}[0]{\overline{\ell}}
\def\mystrut{\vrule height 12pt depth 1pt width 0pt}

\def\simlt{\mathrel{\lower2.5pt\vbox{\lineskip=0pt\baselineskip=0pt
           \hbox{$<$}\hbox{$\sim$}}}}
\def\simgt{\mathrel{\lower2.5pt\vbox{\lineskip=0pt\baselineskip=0pt
           \hbox{$>$}\hbox{$\sim$}}}}

\begin{document}

\setcounter{page}{1}

\begin{titlepage}
\begin{flushright}
ACT-7-20\\ MI-TH-2030
\end{flushright}

\unitlength = 1mm~\\
\vskip .2cm
\begin{center}

{\LARGE{\textsc Cosmology of the string derived flipped $SU(5)$}}

\vspace{0.3cm}
 {\large I. Antoniadis}\,{}\footnote{{\tt antoniad@lpthe.jussieu.fr}}$^{a}$  {\large D.V. Nanopoulos}\,{}\footnote{\tt dimitri@physics.tamu.edu}$^{b}$   {\large J. Rizos}\,{}\footnote{\tt irizos@uoi.gr}${}^c$

\vspace{0.2cm}

{\it  ${}^a$ Laboratoire de Physique Th\'eorique et Hautes Energies - LPTHE\\ Sorbonne Universit\'e, CNRS, 4 Place Jussieu, 75005 Paris, France; \\ 
Albert Einstein Center, Institute for Theoretical Physics, University of Bern,  \\ Sidlerstrasse 5, 3012 Bern, Switzerland\\ 
	  ${}^b$ George P. and Cynthia W. Mitchell Institute for Fundamental Physics and Astronomy, Texas A\&M University, College Station, TX 77843, USA; \\
Astroparticle Physics Group, Houston Advanced Research Center (HARC)\\ Mitchell Campus, Woodlands, TX 77381, USA;\\
Academy of Athens, Division of Natural Sciences, Athens 10679, Greece\\
${}^c$ Physics Department, University of Ioannina, 45110, Ioannina, Greece;\\
School of Science and Technology, Hellenic Open University,
Tsamadou 13-15, GR-26222 Patras, Greece 
\\
}

\vspace{0.4cm}

\begin{abstract}
We study the cosmology of a string derived supersymmetric flipped $SU(5)$ model in the context of free-fermionic heterotic constructions that allow full calculability of the effective supergravity in perturbation theory around the fermionic vacuum where all string moduli have fixed values. The model has 3 generations of chiral families and a Higgs sector leading to particle phenomenology consistent with low energy data, that has been extensively studied in the past. Here, we show that it can also accommodate a novel successful cosmology, based on the no-scale effective supergravity derived from string theory as well as an appropriate induced superpotential suppressed by five powers of the string scale. It utilises two gauge singlet chiral superfields present in the low energy spectrum: the inflaton $y$, identified as the superpartner of a state mixed with R-handed neutrinos, 
and the goldstino $z$ with a superpotential of the form $W_I=M_Iz(y-\lambda y^2)$ (in supergravity units) where $\lambda$ is a dimensionless ${\cal O}\left(1\right)$ parameter and $M_I$ the mass scale of inflation generated at 5th order by the breaking of an anomalous $U(1)_A$ gauge symmetry, characteristic of heterotic string chiral vacua. The resulting scalar potential leads to Starobinsky type inflation.
Our results can be easily generalised to a large class of models with similar properties.
\end{abstract}

\setcounter{footnote}{0}
\vspace{.2cm}
\end{center}

\end{titlepage}

\pagestyle{empty}
\pagestyle{plain}

\pagenumbering{arabic}

\tableofcontents
\bibliographystyle{utphys}

\newpage
\section{Introduction}\label{intro}

If string theory describes nature unifying all fundamental interactions, it should be able to explain all physical phenomena at all scales. During the last three decades, string phenomenology made several steps towards this goal in parallel with the progress made in understanding the foundations of the theory. It started with string constructions of particle physics models to continue with concrete models of cosmology. Very little has been done though in explicit constructions of both at the same time. Usually, models of particle physics focus on spectra extending the Standard Model content and interactions to higher energies, while models of early cosmology focus on generating the inflaton potential following moduli stabilisation, ignoring the details of the actual unification of all interactions.

In this work, we make a step towards a simultaneous description of particle phenomenology and cosmology using the same string derived model\cite{Antoniadis:1989zy} within the framework of free-fermionic formulation of four-dimensional (4d) heterotic superstrings\cite{Antoniadis:1986rn}. Although one of the earliest formalisms of realistic string model building, it has the advantage of full calculability in perturbation theory around a vacuum where all moduli except the dilaton are fixed at the fermionic point, characterised by a set of gauge and discrete symmetries. Instead of constructing a new model from scratch, we consider one of the very first successful constructions of the flipped $SU(5)\times U(1)$~\cite{Antoniadis:1989zy}, that was already shown to give a remarkably good particle phenomenology~\cite{Lopez:1989fb, Rizos:1990xn, Ellis:1990vy, Antoniadis:1991fc} and we study its cosmology. Our analysis can also serve as a proof of concept for a class of 4d string models with similar properties.

Our goal is to identify the inflaton among the gauge singlet massless spectrum of the model and derive an inflationary potential consistent with present observations. We focus on Starobinsky-type inflation~\cite{starobinsky}, which was shown~\cite{starobinsky_sugra, Ellis:2013nxa, kl, fkvp, Antoniadis:2014oya, Kounnas:2014gda} to be naturally embeddable in the framework of $N=1$ no scale supergravity~\cite{Cremmer:1983bf} that is the generic low-energy effective field theory of string models and in particular of fermionic constructions where it is calculable to all orders in the string slope parameter $\alpha'$~\cite{Antoniadis:1987zk, Ferrara:1987tp, Lopez:1994ej}.

Following the proposal of Ref.~\cite{Ellis:2013nxa, Ellis:2018zya, Ellis:2019bmm, Ellis:2020lnc}, we identify the inflaton $y$ with the superpartner of 
a state mixed with Right-handed neutrinos and look for generating a superpotential of the form $W_I=M_Iz(y-\lambda y^2)$, in supergravity units, where $z$ is the goldstino superfield, $M_I$ the inflation scale and $\lambda$ a dimensionless tuneable parameter of order ${\cal O}\left(1\right)$. Since the scale of inflation is about five orders of magnitude lower than the string scale, $W_I$ should be generated at higher (non-renormalisable) order, upon suitable vacuum expectation values (VEVs) of fields that drive the theory ``slightly" away from the free-fermionic point. Such VEVs are generated from an anomalous $U(1)_A$ which becomes massive by absorbing the imaginary part of the dilaton (dual to the universal axion) that transforms under the abelian gauge transformation cancelling the anomaly~\cite{Dine:1987xk}. As a result, the $U(1)_A$ D-term is modified by a dilaton dependent constant proportional to the one loop anomaly, imposing non-trivial VEVs for some charged fields in order to satisfy the D-term vanishing condition. These VEVs should also satisfy the F-flatness supersymmetry conditions and they are naturally lower than the string scale by roughly a loop factor.

In this work, we perform the above described analysis by computing all relevant non-renormalisable superpotential terms up to dimension eight, suppressed by powers of the string scale $M_s^n$ with $n\le 5$. We find a consistent set of VEVs, satisfying all D and F-flatness conditions up to this order, that give rise to the desired inflationary superpotential. It is important to emphasise that this set of VEVs describe the inflationary phase of the model, where in particular $SU(5)\times U(1)$ remains unbroken, and not the vacuum after the end of inflation. Indeed, the $SU(5)\times U(1)$ breaking occurs via a first order phase transition at a critical temperature which is lower than the inflation scale~\cite{Ellis:2018moe}. As a result, the previous investigations of the particle phenomenology of the model remain in principle valid, since they describe the low energy vacuum.

The outline of the paper is the following. In Section~\ref{revamped}, we recall for convenience of the reader, the string construction of the ``revamped" flipped $SU(5)\times U(1)$ model~\cite{Antoniadis:1989zy}, its spectrum, the trilinear superpotential and its main properties. In Section~\ref{staroinflation}, we present a review of the Starobinsky-type inflation within the context of no scale supergravity, in the particular form emerging at the low energy limit of 4d free-fermionic heterotic superstrings. In Section~\ref{realisation}, we perform the computation and the analysis of the non-renormalisable F-terms up to eighth order, we identify the inflaton and the goldstino superfield and derive the desired inflationary superpotential. We conclude in Section 5 with a summary of  our results and outlook. Finally, Appendix~A contains the technical details of the string construction, while in Appendix~B we give some useful details of the calculation of the effective field theory for inflation that we discuss in Section~\ref{staroinflation}.

\section{The ``revamped" flipped $SU(5)\times U(1)$ string model}
\label{revamped}

In this section, we give a short review of the flipped $SU(5)$ model~\cite{Antoniadis:1989zy} constructed within the free-fermionic formulation of four-dimensional (4d) heterotic strings~\cite{Antoniadis:1986rn}. For convenience of the reader, we use the same notation as in the original paper~\cite{Antoniadis:1989zy}.
The model is generated by a set of basis vectors of boundary conditions for all world-sheet fermions with an appropriate choice of mutual GSO projections, given for self-consistency in Appendix~A. It is equivalent to a $Z_2\times Z_2$ asymmetric orbifold of a $N=4$ supersymmetric model with gauge group $U(8)\times SO(2)^6\times SO(10)\times SO(6)$, generated by the vectors $\{ S, 1, \zeta, \alpha, b_1+b_4, b_2+b_5\}$. The $Z_2\times Z_2$ operation defined by the vectors $\{ b_1,b_2\}$ breaks supersymmetry to $N=1$, $U(8)\to U(5)\times U(1)^3$ and $SO(2)^6\to SO(2)$. 

Thus, besides $SU(5)\times U(1)$, the gauge group contains a ``hidden sector" $SO(10)\times SO(6)$ and four abelian factors $U(1)^4$, one combination of which has an anomaly cancelled by the Dine-Seiberg-Witten mechanism~\cite{Dine:1987xk} and acquires a mass upon absorbing the universal axion whose shift symmetry is gauged under the anomalous $U(1)_A$ transformation. 
The matter spectrum consists of: 
\begin{itemize}
\item Four chiral and one anti-chiral generations with the quantum numbers of quarks and leptons, denoted as $(F_i, \of_i, \ell^c_i)$ transforming under $SU(5)$ as $({\bf 10},{\bf\bar 5},{\bf 1})$ respectively and forming $SO(10)$ spinors; two are coming from the first twisted plane $(i=1, 4)$ in the sectors $b_1$ and $b_4$, one from the third twisted plane $(i=3)$ in the sector $b_3$, while the second twisted plane gives a pair of a chiral and anti-chiral generations $(i=2,5)$ from the sectors $b_2$ and $b_5$, respectively.
\item Four pairs of $({\bf 5}+{\bf\bar 5})$ forming $SO(10)$ vectors and containing electroweak higgses; three pairs are coming from the Neveu-Schwarz (NS) $0$-sector $(h_i, {\bar h}_i)_{i=1,2,3}$ and one pair from the 3rd twisted plane $(h_{45}, {\bar h}_{45})$ from the sector $S+b_4+b_5$.
\item Five vector representations $({\bf 10},{\bf 1})+({\bf 1},{\bf 6})$ of the hidden gauge group $SO(10)\times SO(6)$, denoted as $T_i$ and $D_i$, respectively, in the sectors $b_i+2\alpha+\zeta$ and $b_i+2\alpha$; thus, two are coming from the first twisted plane $(i=1, 4)$, two from the second $(i=2, 5)$, and one from the third $i=3$.
\item Six pairs of $SO(6)$ spinors $({\bf 4}+{\bf\bar 4})$ with fractional electric charges $\pm 1/2$ that confine into integrally charged bound states; in their notation below, $X,Y,Z$ correspond to the 1st, 2nd and 3rd twisted plane, correspondingly.
\item Ten pairs of singlets under the observable and hidden gauge groups, but charged under the four $U(1)$'s; seven are coming from the third twisted sector $S+b_4+b_5$: $\phi_1,\dots,\phi_4$, $\phi_\pm$, $\phi_{45}$ (and their conjugates $\bar\phi_i$, etc), while three are coming from the NS sector: $\Phi_{23},\Phi_{31},\Phi_{12}$ (and their conjugates).
\item Five gauge singlets from the NS sector: $\Phi_1,\dots,\Phi_5$.
\end{itemize}
The complete spectrum and quantum numbers are presented in Appendix A.
The full tree-level superpotential reads
\begin{align}
W_3 =\, 
& g_s \sqrt{2}\left[\mystrut F_1 F_1 h_1 + F_2 F_2 h_2 + F_4 F_4 h_1 + \oF_5 \oF_5 \oh_2 
+F_4 \of_5 \oh_{45}+F_3 \of_3 \oh_3\right.\nonumber\\
&
+\of_1 \ell^c_1 h_1+\of_2 \ell^c_2 h_2 +\of_5 \ell^c_5 h_2  + f_4 \oell^c_4 \oh_1\nonumber\\
& +
\frac{1}{\sqrt{2}} F_4 \oF_5 \phi_3 + \frac{1}{\sqrt{2}} f_4 \of_5 \ophi_2 + \frac{1}{\sqrt{2}} \oell^c_4 \ell^c_5 \ophi_2 
\nonumber\\
& + h_1 \oh_2 \Phi_{12} + \oh_1 h_2 \oPhi_{12} + h_2\oh_3\Phi_{23} + \oh_2 h_3 \oPhi_{23}
+ h_3 \oh_1\Phi_{31} \nonumber\\ 
&+ \oh_3 h_1\oPhi_{31}+h_3 \oh_{45}\ophi_{45} + \oh_3 h_{45} \phi_{45}
+ \frac{1}{2} h_{45} \oh_{45} \Phi_3 \nonumber\\
&+\phi_1\ophi_2 \Phi_4 + \ophi_1\phi_2 \Phi_4+  \phi_3 \ophi_4 \Phi_5 + \ophi_3 \phi_4 \Phi_5
+\frac{1}{2}\phi_{45} \ophi_{45} \Phi_3\nonumber\\
&+\frac{1}{2}\phi_{+} \ophi_{+} \Phi_3 
+\frac{1}{2}\phi_{-} \ophi_{-} \Phi_3 \label{wtree}\\
&+\Phi_{12} \Phi_{23} \Phi_{31} +\oPhi_{12} \oPhi_{23} \oPhi_{31}  +
\Phi_{12} \phi_+ \phi_- + \oPhi_{12} \ophi_+ \ophi_- 
\nonumber\\
&+
\frac{1}{2} \Phi_3 \sum_{i=1}^4 \phi_i \ophi_i +  \Phi_{12} \sum_{i=1}^4 \phi_i^2 
+  \oPhi_{12} \sum_{i=1}^4 \ophi_i^2\nonumber\\
&+D_1^2\oPhi_{23}+D_2^2\Phi_{31}+D_4^2\oPhi_{23}+D_5^2\oPhi_{31}+\frac{1}{\sqrt{2}} D_4 D_5 \ophi_3 \nonumber\\
&+T_1^2\oPhi_{23}+T_2^2\Phi_{31}+T_4^2\Phi_{23}+T_5^2\Phi_{31}+\frac{1}{\sqrt{2}} T_4 T_5 \phi_2 \nonumber\\
&+\frac{1}{\sqrt{2}} Y_1 \overline{X}_2 \phi_4 +\frac{1}{\sqrt{2}} Y_2 \overline{X}_1 \phi_1 + Y_2 \overline{X}_2\phi_+ 
+ \frac{1}{2} Z_1 \overline{Z}_1 \Phi_3\nonumber\\
&\left.+Z_2 \overline{Z}_2 \oPhi_{12}+Z_1 \overline{X}_2' \ell^c_2 + Y_2' Z_1 D_1\mystrut\right]\,,\nonumber
\end{align}
where $g_s$ is the string coupling.
The  breaking of the Grand Unified Theory (GUT) group to the Standard Model (SM) is achieved by the VEVs $\langle{\bar F}_5\rangle=\langle F\rangle$, where $F\equiv\sum_{i=1}^3\alpha_i F_i$, along a supersymmetric F-flat direction that leaves the coefficients $\alpha_i$ of the linear combination $F$ undetermined \cite{Antoniadis:1989zy}. These are accompanied by a set of VEVs for $SU(5)\times U(1)$ singlets subject to F- and D-term vanishing conditions, where in the former higher order (non-renormalisable) terms play in principle an important r\^ole.
The phenomenology of the model, including quark masses and mixings, baryon decay and neutrino masses  has been studied extensively in the past 
\cite{Antoniadis:1989zy, Lopez:1989fb, Rizos:1990xn, Ellis:1990vy, Antoniadis:1991fc}. 
The effective field theory of the model is a $N=1$ no-scale supergravity~\cite{Cremmer:1983bf} with a K\"ahler potential of chiral multiplets that can be computed exactly at the string tree-level (to all orders in $\alpha'$) using the method of Refs.~\cite{Antoniadis:1987zk, Ferrara:1987tp, Lopez:1994ej}, applied for generic 4d free-fermionic heterotic superstrings.

\section{Starobinsky type inflation from no-scale supergravity}
\label{staroinflation}

It is shown that Starobinsky type inflation~\cite{starobinsky} can be easily embedded in no-scale supergravity implemented with an appropriate superpotential for the inflaton field~\cite{starobinsky_sugra,  Ellis:2013nxa, kl, fkvp, Antoniadis:2014oya, Kounnas:2014gda}. Here we will use a particular proposal~\cite{Ellis:2013nxa, Ellis:2018zya, Ellis:2019bmm, Ellis:2020lnc} that can be naturally derived in the effective field theory of free-fermionic heterotic constructions. 

The Starobinsky model of inflation is obtained by adding to the Einstein action a quadratic in the scalar curvature term. This can be linearised by introducing a Lagrange multiplier which becomes, in the Einstein frame, an ordinary scalar field (the so-called scalaron) with a particular potential~\cite{Whitt:1984pd} that can drive inflation compatible with present cosmological observations.
Since the scalar curvature belongs in supergravity to a chiral multiplet~\cite{SUGRA}, the supersymmetric extension of $R^2$ must be written as a D-term, which needs two superfield-Lagrange multipliers to be linearised~\cite{cecotti}. One of them contains the inflaton, while the other is the goldstino superfield that breaks supersymmetry. The model is described by a no-scale type supergravity with a K\"ahler potential $K$ and a superpotential $W$ given by:
\be
\label{staro}
K=-3\ln \left(T+{\bar T} - C{\bar C}\right)\quad;\quad W=\mu C(T-{k})\,,
\ee
where $\mu$ is the inflation scale and $k$ is an irrelevant constant that can be set to 1 by an appropriate rescaling of the fields ($T$ by $k$ and $C$ by $k^{1/2}$). The normalised inflaton $\varphi$ is given by ${\rm Re} T=e^{\sqrt{2/3}\varphi}$, while the linear term in $C$ of the superpotential drives the supersymmetry breaking.

Since in the free-fermionic constructions all moduli (except the dilaton) are fixed to particular values in general of enhanced symmetries (gauge or discrete), it is convenient to go to an appropriate basis where fields have vanishing VEVs, describing fluctuations around the free-fermionic point. Defining
\be
\label{transfo}
T={1+y\over 1-y}\quad;\quad  C=\sqrt{2}{z\over 1-y}\,,
\ee
one obtains:
\be
\label{staro2}
K=-3\ln \left(1 - |y|^2-|z|^2\right)\quad;\quad W=\mu z\left(y-y^2\right)\,,
\ee
upon a suitable K\"ahler transformation.
The inflation region corresponds to $\varphi\to\infty$, or equivalently $|y|\to 1$, for vanishing $C$ or $z$, where the scalar potential is approximately constant $\propto\mu^2$. The main problem of this model is that the sgoldstino $C$ (or $z$) is unstable (tachyonic) during inflation~\cite{kl}. Stabilisation of the runaway direction can be achieved by adding for instance a quartic (or higher order) term in the K\"ahler potential which obviously modifies the theory beyond the minimal supersymmetrisation of $R^2$. Alternatively, one can parametrise the stabilisation mechanism in a model independent way by integrating out the heavy sgoldstino replacing $C$ with the nilpotent goldstino superfield satisfying the constraint $C^2=0$~\cite{Antoniadis:2014oya}. Here, we will follow a different approach based on the {\em derived} effective field theory of free-fermionic string models.

The first observation is that the K\"ahler potential is in general a sum of different factors, splitting the logarithm \eqref{staro2} into pieces. When twisted fields are set to zero, the K\"ahler potential of all untwisted fields splits in four pieces:
\bea
\label{KUT}
K_{UT}=K_0+\sum_{i=1}^3K_i\quad;\quad K_0 &=& -\ln(S+{\bar S})\\ 
\,,
K_i &=& -\ln\left(1-\sum_{A=1}^{n_i} \left|\Phi_{i,A}\right|^2+{1\over 4}\left|\sum_{A=1}^{n_i}\Phi_{i,A}^2\right|^2\right) 
\nonumber
\eea
where $S$ denotes the dilaton superfield and $K_i$ parametrise an $SO(2,n_i)/SO(2)\times SO(n_i)$ manifold~\cite{Antoniadis:1987zk, Ferrara:1987tp}. As we mentioned in the previous section, a $N=1$ supersymmetric free-fermionic construction corresponds to a $Z_2\times Z_2$ asymmetric orbifold acting on a $N=4$ theory. The orbifold group has therefore three $Z_2$ non-trivial elements $h_i$, $i=1,2,3$. The action of every element alone gives rise to a $N=2$ theory. The set of the $n_i$ chiral superfields $\{\Phi_{i,A}\}$ corresponds to those coming from the vector multiplets of the $i$-th $N=2$ theory surviving the other two projections. In the weak field limit, twisted fields give an additional contribution to the K\"ahler potential \eqref{KUT}:
\be
\label{KT}
K_T=\sum_{i=1}^3 K_{T_i}\quad;\quad K_{T_i}\simeq \sum_{\{\Phi_{T_i}\}} |\Phi_{T_i}|^2 e^{K_j+K_l}\,,\, i\ne j\ne l\ne i\,,
\ee
where the sum is extended over all chiral superfields of the $i$-th twisted sector. On the other hand, in the case one keeps the fields of a single twisted sector, say $T_1$, $K_{T_1}$ becomes to all orders~\cite{Ferrara:1987tp}:
\be
\label{KT1}
K_{T_1}= -2\ln\left(1-{1\over 2} \sum_{\{\Phi_{T_1}\}} |\Phi_{T_1}|^2 e^{{1\over 2}(K_2+K_3)} \right)\,.
\ee
Adding $K_2+K_3$ from \eqref{KUT} to $K_{T_1}$, one thus has
\be
\label{KT2}
K_2+K_3+K_{T_1}= -2\ln\left( e^{-{1\over 2}(K_2+K_3)}-{1\over 2} \sum_{\{\Phi_{T_1}\}} |\Phi_{T_1}|^2 \right)\,.
\ee

In the proposal~\cite{Ellis:2018moe}, the inflaton $y$ was identified with the superpartner of a state that mixes with a R-handed neutrino 
entering in the so-called $\lambda_6$ coupling of the flipped $SU(5)$ superpotential~\cite{Antoniadis:1989zy} that provides neutrino masses upon the GUT symmetry breaking. The inflaton can then decay in R-handed neutrinos reheating the universe and there is an interesting connection with the neutrino mass generation. This possibility is compatible with the GUT symmetry breaking occurring by a first order transition at a critical temperature which is lower than the inflation scale \cite{Ellis:2018moe}. As a result, $SU(5)\times U(1)$ remains unbroken during the inflationary era.

In the model described in the previous section, the $\lambda_6$ coupling corresponds to the term $F_4{\bar F}_5\phi_3$ in the superpotential \eqref{wtree}. The inflaton $y$ should then correspond to a linear combination of fields containing $\phi_3$ that originates from the third twisted sector $T_3$. By inspection of the K\"ahler potentials \eqref{KUT} - \eqref{KT2}, we then consider the possibility that the goldstino superfield $z$ appearing in the inflationary superpotential \eqref{staro} comes from the third untwisted sector, so that its corresponding K\"ahler potential $K_3$ does not appear inside the logarithm \eqref{KT1}. Using $K_3$ from \eqref{KUT} for the case of one field $z$ parametrising an $SO(2,1)/SO(2)$ manifold and restricting \eqref{KT2} to 
a single field $y\equiv\Phi_{T_3}/\sqrt{2}$, one obtains:
\begin{align}
\label{staro3}
K&=-\ln(S+{\bar S})-2\ln\left(1-\frac{1}{2}|z|^2\right)-2\ln\left(1-\left|y\right|^2\right)\\
W&= \mu z \left(y- y^2\right) \nonumber
\end{align}
where all other fields are set to zero, corresponding to the fermionic point; VEVs away from it are treated perturbatively. 

Defining $y=\rho e^{i\theta}$ with $0\le\rho<1$, the resulting scalar potential reads
\footnote{For details of the derivation see Appendix B.}:
\begin{align}
\label{scalar-potential}
V &\!=\! e^{K} \left(\sum_{\Phi=S,y,z}\left|D_\Phi W\right|^2-3\left|W\right|^2\right) \\
&\!=\! \mu^2 g_s^2\!\left\{\!
\frac{4 \rho^2+2(1\!-\!2\rho\!-\!\rho^2)\left|z\right|^2+\rho^2\left|z\right|^4}{(1+\rho)^2\left(2-\left|z\right|^2\right)^2}\! +
\! 4\frac{4\rho^2+4|z|^2+\rho^2|z|^4}{\left(1-\rho^2\right)^2\left(2-\left|z\right|^2\right)^2} \rho\sin^2{\theta\over 2}
\!\right\}, \nonumber
\end{align}
where $D_\Phi\equiv\partial_\Phi +(\partial_\Phi K)$ is the K\"ahler covariant derivative and
$g_s$ is the string coupling which we consider constant assuming a suitable stabilisation mechanism for the dilaton. 
Minimisation with respect to the phase gives $\theta=0$ and thus the potential becomes:
\be
\label{Vtheta0}
V{\Large\mid}_{\theta =0}= \frac{\mu^2 g_s^2}{(1+\rho)^2}\left(\rho^2+\frac{1}{2}\frac{(1-\rho)^2\left|z\right|^2}{\left(1-\frac{\left|z\right|^2}{2}\right)^2}\right)\;.
\ee
It is easy to see that the phase fluctuation acquires a mass of order the inflation scale and decouples during inflation.
We also find very interesting that the sgoldstino direction $z$ is not any more unstable during inflation. The dilaton contribution raised the tachyonic mass to a positive value so that $z$ becomes asymptotically a flat direction in the limit $\rho\to 1$. More precisely, in the absence of the dilaton, the factor $(1-\rho)^2$ multiplying $|z|^2$ in the expression \eqref{Vtheta0} becomes $(1-\rho)^2-2\rho^2$. Note that the mass of $z$ is proportional to the inflaton slope, leading to possible non-gaussianities which require a separate analysis.

In order to study the supersymmetry breaking, one should compute the values of the auxiliary fields 
$F_\Phi=e^{K/2}D_\Phi W$ for all fields $\Phi$. It is easy to see that all of them vanish at $\theta=z=0$ except for $F_z$:
\be
\label{susy}
F_y{\Large\mid}_{z=\theta =0}=F_S{\Large\mid}_{z=\theta =0}=0\quad;\quad 
F_z{\Large\mid}_{z=\theta =0}=\mu g_s\frac{\rho}{1+\rho}\,.
\ee
It follows that, as expected, $z$ plays the r\^ole of the goldstino superfield that breaks spontaneously supersymmetry during inflation at a scale $\mu g_s/2$. On the other hand, in the absence of soft breaking terms that play no role in the inflation mechanism, the minimum of the potential is supersymmetric.

The canonically normalised inflaton field $\varphi$ is given by:
\be
\label{phi-rho}
\varphi = \ln\frac{1+\rho}{1-\rho}\quad;\quad \rho=\tanh{\varphi\over 2}\,,
\ee
and the potential at $z=0$ becomes
\be
\label{Vztheta0}
V{\Large\mid}_{z=\theta =0}= \mu^2 g_s^2\left( \frac{\rho}{1+\rho}\right)^2
=\frac{\mu^2 g_s^2}{4}\left( 1-e^{-\varphi}\right)^2\,.
\ee
For details see Appendix B.
It has the same form as in the case $\alpha=2/3$ of Ref.~\cite{Ellis:2019bmm} from which one can deduce the inflationary predictions of the model for the spectral index $n_s$ of primordial density perturbations and the ratio $r$ of tensor-to-scalar modes:
\be
\label{predictions}
n_s=0.965\quad;\quad r=2.4\times 10^{-3}\,,
\ee
for the canonical value of about $N_\ast=55$ e-folds of inflation.
At the minimum of the potential, $\varphi=0$, the inflaton gets a mass $m_\varphi=\mu g_s/\sqrt{2}$.

In the next section, we  derive the model we described above in string theory within the ``revamped" flipped $SU(5)\times {U(1)}_A$ construction we described in Section~\ref{revamped}. As we discussed already, the K\"ahler potential \eqref{staro3}  is fixed by an appropriate choice of fields and the main goal of the analysis will be to compute the inflationary superpotential. Our strategy is to generate it at a higher non-renormalisable level, via VEVs of $SU(5)\times U(1)$ singlet fields, since this gauge symmetry remains unbroken at the inflation scale. Non-trivial VEVs are driven by the anomalous $U(1)_{A}$ cancellation mechanism and are naturally a loop factor below the string scale. One may also worry on the relative coefficient ($-1$ in our basis) between the quadratic and cubic term in $W$, in (supergravity) Planck units, implying in general a fine-tuning of the corresponding VEVs. Replacing $-1$ by an order one parameter $-\lambda$:
\be
\label{Wlambda}
W= \mu z \left(y-\lambda y^2\right) 
\ee
(the sign is not relevant since it can change by redefining $y\to -y$), the potential will diverge at the boundary $y=1$ but an inflationary period of about 60 e-folds remains present for a range of values of $\lambda$ around unity within about $10^{-3}$~\cite{Ellis:2020lnc}.\footnote{Note the difference between our model and the one quoted in Ref.~\cite{starobinsky_sugra,Ellis:2013nxa,Ellis:2020lnc}.}

\section{Realisation of a Starobinsky-like inflationary scenario in the flipped ${SU(5)}\times{U(1)}$ string model}
\label{realisation}
In this section we consider a concrete realisation of the Starobinsky-like inflationary scenario described in Section 3 in the context of the ``revamped" ${SU(5)}\times{U(1)}$ string model presented in Section 2. 
In this framework, a central r\^ole is played by the so called $\lambda_6$ superpotential coupling, that is the coupling relating a gauge singlet, the GUT breaking Higgs and matter fields, which eventually provides the neutrino see-saw mechanism.
In the flipped $SU(5)$ string model under consideration there exists a single term of this type in the tree-level superpotential \eqref{wtree}, namely $F_4 \overline{F}_5 \phi_3$, where $F_4$ is a matter field and $\overline{F}_5$ is the flipped $SU(5)$ breaking Higgs field. Following Section 3 the singlet $\phi_3$ has to be linked  to the inflaton field. This in turn requires $\phi_3$ to remain massless at leading order in the superpotential, whereas it may acquire Majorana mass and mixings with other singlet field(s) at some higher order. 
As a matter of fact, intermediate mass scales can be generated in the context of low energy string effective theories via the use of non-renormalisable interactions, as follows.
In the presence of an anomalous abelian symmetry D-flatness requires some fields in the theory to develop VEVs one or two orders of magnitude below the Planck scale. 
These in turn can entail additional field VEVs via the F-flatness conditions. 

A superpotential term of the form $\phi_1\phi_2\phi_3\varphi^{N-3}$, where $\varphi^{N-3}$ stands for a product of $N-3$ fields acquiring VEVs, gives rise to an effective interaction $\lambda \phi_1\phi_2\phi_3$, with 
$\lambda\sim\left(\frac{\langle\varphi\rangle}{M_P}\right)^{N-3}$. However, these non-renormalisable terms are subject to non-trivial string selection rules that go beyond mere gauge group invariance \cite{Lopez:1990wt,Rizos:1991bm}. For example, a pure Neveu–Schwarz (NS) coupling, that is a coupling consisting exclusively of fields arising from the $0/S$ sector (vectors of boundary conditions), 
vanishes identically for any $N>3$. More particularly, it can be shown, that in the model under consideration, all non-renormalisable interactions involving solely singlet fields vanish identically \cite{Lopez:1990wt}. 
This stems from the fact that in the specific model non-abelian gauge group singlets come solely from two sectors, the NS $0/S$ sector and the third twisted sector $T_3$ of the $b_4+b_5/S+b_4+b_5$ vectors of boundary conditions.
It follows that generation of a non-trivial bilinear or trilinear superpotential involving $\phi_3$ at a higher order, requires the introduction of additional VEVs for some hidden sector fields (twisted in the first and/or the second plane). In view of this requirement, and in order to examine the feasibility of Starobinsky-like inflationary scenario, we start by deriving the constraints on all possible VEV assignments for $SU(5)\times U(1)$ gauge singlet fields that may transform non-trivially under the hidden group $SO(10)\times SO(6)$. 


The revamped flipped string model gauge symmetry is $[SU(5)\times{U(1)}]\times{U(1)}^4\times{[SO(10)}\times{SU(4)]}$.
As explained in Appendix A, the abelian factor product ${U(1)}^4$ can be recast in the form $\left\{\prod_{I=1}^3{U(1)}_I'\right\}\times{U(1)}'_A$, where $U(1)'_I, I=1,2,3$ are anomaly free and $U(1)'_A$ is anomalous. The anomalous abelian symmetry is broken via the Dine-Seiberg-Witten mechanism at the loop level \cite{Dine:1987xk}, involving therefore a VEV of the dilaton field, {\em i.e.} the string coupling. The induced Fayet-Iliopoulos term breaks supersymmetry unless it is cancelled by a VEV along 
an appropriate matter field combination.
In this case, the D-flatness conditions associated with the ${U(1)'}^4$ group factor read~\cite{Antoniadis:1989zy, Cvetic:1998gv}:
\begin{align}
{\cal D}_I &= \sum_{i} q^i_I \left|\varphi_i\right|^2=0\,, I=1,2,3\,,\label{DFI}\\
{\cal D}_A &= \sum_{i} q^i_A \left|\varphi_i\right|^2+
\frac{1}{192\pi^2}{2\over\alpha'}\,{\rm Tr}Q_A=0\,,
\label{DFA}
\end{align}
where 
$\varphi_i$ stand for massless fields with charges $q^i_I/q^i_A$ under ${U(1)'}_I/{U(1)'_A}$
and the trace of the anomalous $U(1)'_A$ generator ${\rm Tr}{Q_A}=180$. 
In addition, $N=1$ supersymmetry imposes the F-flatness constraints
\begin{align}
{\cal F}_i = \frac{\partial W}{\partial\varphi_i}=0\,, \label{FFC}
\end{align} 
where $W$ stands for the superpotential.

For our purposes, we will assume VEVs for the singlet fields $\Phi_A,\oPhi_A,A=12,23,31\,$, $\Phi_i,i=1,\dots,5,\, \phi_j,\ophi_j, j=1,2,3,4,+,-,45$ as well as the 
``hidden" sector fields $D_a, T_a\,a=1,\dots,5$.
Following \cite{Rizos:1990xn}, the 
D-flatness conditions \eqref{DFI} and \eqref{DFA} can be recast in the form
\footnote{$D_a$/$T_a, a=1,\dots,5$ fields carry also group indices as they transform in the vectorial representation of $SO(6)/SO(10)$ gauge group factor respectively. In the sequel, we suppress group indices and employ a dot notation to indicate group index contraction, e.g. $D_a\cdot{D_b}$
stands for $\sum_i D_a^i{D_b}^i$. Group index contraction is also implied 
in expressions like $T_a^2\equiv T_a{\cdot}T_a$, $\left|T_a\right|^2\equiv T_a{\cdot}T_a^*$.}:
\begin{align}
&\left|\phi_{45}\right|^2  - \left|\overline{\phi}_{45}\right|^2 - \frac{1}{2}\left( \left|D_3\right|^2 + \left|T_3\right|^2\right) = 
\xi^2\,, \label{df1}\\
&\left|\phi_+\right|^2 - \left|\phi_-\right|^2 - \left|\overline{\phi}_+\right|^2 + \left|\overline{\phi}_-\right|^2 + \frac{1}{2}
\left(\left|D_3\right|^2  - \left|T_3\right|^2\right) = \xi^2\,,\label{df2}\\
&\left|\Phi_{31}\right|^2 -\left|\overline{\Phi}_{31}\right|^2 
-\left|\Phi_{23}\right|^2 +\left|\overline{\Phi}_{23}\right|^2 
-\frac{1}{2}\left(\left|D_1\right|^2 + \left|D_2\right|^2 + \left|D_3\right|^2 + \left|D_4\right|^2 - \left|D_5\right|^2\right)\nonumber\\
&\ 
-\frac{1}{2} \left(\left|T_1\right|^2 + \left|T_2\right|^2 + \left|T_3\right|^2 - \left|T_4\right|^2 + \left|T_5\right|^2\right)
= 3\xi^2\,,\label{df3}
\end{align}
\begin{align}
&\left|\Phi_{23}\right|^2  - \left|\overline{\Phi}_{23}\right|^2
-\left|\Phi_{12}\right|^2  + \left|\overline{\Phi}_{12}\right|^2
+\frac{1}{2}\sum_{i=1}^4 \left(\left|\phi_i\right|^2-\left|\overline{\phi}_i\right|^2\right)
+ \left|\phi_+\right|^2 - \left|\overline{\phi}_+\right|^2 \nonumber\\
&\ +\frac{1}{2}\left(\left|D_1\right|^2 + \left|D_3\right|^2 + \left|D_4\right|^2\right)
+ \frac{1}{2}\left(\left|T_1\right|^2 - \left|T_4\right|^2 \right) = 0\,,
\label{df4}
\end{align}
where
\begin{align}
\xi^2 =\frac{1}{16\pi^2}{2\over\alpha'}\,.
\label{xidef}
\end{align}
The F-flatness conditions can be computed utilizing \eqref{FFC}. Restricting to the tree-level superpotential contributions \eqref{wtree} we have
\begin{align}
\left({\Phi_{12}}\right) &: \sum_{i=1}^4\phi_i^2+\Phi_{23}\Phi_{31} + \phi_+ \phi_-=0\,,
\label{firste}\\
\left(\overline{\Phi}_{12}\right) &: \sum_{i=1}^4\overline{\phi}_i^2+\overline{\Phi}_{23}\overline{\Phi}_{31}
+ \overline{\phi}_+ \overline{\phi}_-=0\,,\\
\left(\Phi_{31}\right) &: \Phi_{12} \Phi_{23} +  D_2^2 + T_2^2 +T_5^2 =0\,,\\
\overline{\Phi}_{31} &: \overline{\Phi}_{12}\overline{\Phi}_{23} +  D_5^2 =0\,,\\
\left({\Phi}_{23}\right) &: {\Phi}_{12}{\Phi}_{31}+ T_4^2 =0\,,\\
\left(\overline{\Phi}_{23}\right) &: \overline{\Phi}_{12}\overline{\Phi}_{31} + D_1^2+D_4^2+T_1^2 =0\,,\\
\left(\Phi_3 \right) &: \sum_{i=1}^4\phi_i\overline{\phi}_i + \phi_{45}\overline{\phi}_{45} + 
\phi_-\overline{\phi}_- + \phi_+\overline{\phi}_+ =0\,,\\
\left(\Phi_4\right)  &: \phi_2\overline{\phi}_1 + \phi_1\overline{\phi}_2 = 0\,,\\
\left(\Phi_5\right)  &: \phi_4\overline{\phi}_3 + \phi_3\overline{\phi}_4 = 0\,,
\end{align}
\begin{align}
\left(\phi_1 \right)&: 2 \phi_1 \Phi_{12} + \Phi_3\overline{\phi}_1 + \Phi_4 \overline{\phi}_2 = 0\,,\\
\left(\phi_2\right) &:  2 \phi_2 \Phi_{12} + \Phi_4\overline{\phi}_1 + \Phi_3 \overline{\phi}_2  + T_4{\cdot}T_5 = 0\,,\\
\left(\phi_3\right) &: 2 \phi_3 \Phi_{12} + \Phi_3\overline{\phi}_3 + \Phi_5 \overline{\phi}_4  = 0\,,\\
\left(\phi_4\right) &: 2\phi_4 \Phi_{12} + \Phi_5\overline{\phi}_3 + \Phi_3 \overline{\phi}_4 = 0\,,\\
\left(\overline{\phi}_1 \right)&: 2 \ophi_1 \oPhi_{12} + \Phi_3 {\phi}_1 + \Phi_4 {\phi}_2 = 0\,,\\
\left(\ophi_2\right) &:  2 \ophi_2 \oPhi_{12} + \Phi_4 {\phi}_1 + \Phi_3{\phi}_2   = 0\,,\\
\left(\ophi_3\right) &: 2 \ophi_3 \oPhi_{12} + \Phi_3 {\phi}_3 + \Phi_5 {\phi}_4  + D_4{\cdot}D_5= 0\,,
\label{d45}\\
\left(\ophi_4\right) &: 2\ophi_4 \oPhi_{12} + \Phi_5 {\phi}_3 + \Phi_3 {\phi}_4 = 0\,,\\
\left(\phi_+\right) &: \Phi_3 \ophi_+ + \Phi_{12} \phi_- = 0\,, \label{ep}\\
\left(\ophi_+\right) &: \Phi_3 \phi_+ + \oPhi_{12} \ophi_- = 0\,,\label{epb}\\
\left(\phi_-\right) &: \Phi_3 \ophi_- + \Phi_{12} \phi_+ = 0\,,\label{em}\\
\left(\ophi_-\right) &: \Phi_3 \phi_- + \oPhi_{12} \ophi_+ = 0\,,\label{emb}\\
\left(\phi_{45}\right) &: \Phi_3 \ophi_{45} = 0 \,,\label{p45}\\
\left(\ophi_{45} \right)&: \Phi_3 \phi_{45} = 0\,, \label{p45b}
\\
\left(T_1\right) &: 2 \overline{\Phi}_{23} T_1 =0\,,\\
\left(T_2\right) &: 2 \Phi_{31} T_2 =0\,,\\
\left(T_4\right) &: 2 \Phi_{23} T_4 + \phi_2 T_5 =0\,,\\
\left(T_5\right) &:  \phi_2 T_4 + 2 \Phi_{31} T_5 = 0\,,\\
\left(D_1\right) &: 2 D_1 \overline{\Phi}_{23} = 0\,,
\label{d1f}\\
\left(D_2 \right)&: 2 D_2 {\Phi}_{31} = 0\,,\\
\left(D_4\right) &: 2 D_4 \overline{\Phi}_{23} + D_5 \overline{\phi}_3 = 0\,,
\label{d4f}\\
\left(D_5\right) &: 2 D_5 \overline{\Phi}_{31} + D_4 \overline{\phi}_3 = 0\,,\hspace{3cm}
\label{laste}
\end{align}
where $(\varphi_i)$ denotes the flatness equation corresponding to  ${\partial W}/{\partial\varphi_i}$.

A solution of the F/D-flatness equations \eqref{firste}-\eqref{laste} involves, in general,  a set of fields with vanishing VEVs and a set of fields with VEVs of order $\xi$.
For the purpose of our study and after trial and error analysis, we will focus on the subspace of solutions defined by
\begin{gather}
\Phi_{12}=\oPhi_{12} = \Phi_{23}=\oPhi_{23} = \oPhi_{31} = 0\,,\nonumber\\
\Phi_3=\Phi_4=\Phi_5=0\,,\nonumber\\
\phi_1=\phi_2=\phi_3=\ophi_1=\ophi_2=\ophi_3=0\,,\label{flata}\\
\oF_5=F_1=F_2=F_3=F_4=0\,,\nonumber\\
D_2 = T_2 = T_5 = 0\,,\nonumber\\
T_4^2 = D_5^2 = D_4{\cdot}D_5 = 0\,.\nonumber
\end{gather}
Note that $T_a$/$D_a$ are complex $SO(10)$/$SO(6)$ vectors, and thus, they can be nilpotent without being vanishing.
The above choice has the advantage of reducing the tree-level F-flatness equations \eqref{firste}-\eqref{laste} to
\begin{align}
(\Phi_{12}) &: \phi_4^2 + \phi_+ \phi_- =0\,,\label{etf}\\
(\oPhi_{12}) &: \ophi_4^2 + \ophi_+ \ophi_- =0\,,\\
\left(\oPhi_{23}\right) &: D_1^2 + D_4^2 + T_1^2 = 0\,,\\
\left(\Phi_3\right) &: \phi_4\ophi_4 + \phi_{45} \ophi_{45} + \phi_+\ophi_+ + \phi_-\ophi_- = 0\,.
\label{etl}
\end{align}
In addition, the D-flatness conditions \eqref{df1}-\eqref{df2} remain unchanged while \eqref{df3}-\eqref{df4} now read
\begin{align}
	&\left|\Phi_{31}\right|^2 
	-\frac{1}{2}\left(\left|D_1\right|^2  + \left|D_3\right|^2 + \left|D_4\right|^2 - \left|D_5\right|^2\right)
	-\frac{1}{2} \left(\left|T_1\right|^2 +  \left|T_3\right|^2 - \left|T_4\right|^2 \right)
	= 3\xi^2\,,\label{ddf3}\\
	&
	\left|\phi_4\right|^2-\left|\overline{\phi}_4\right|^2
	+ 2\left(\left|\phi_+\right|^2 - \left|\overline{\phi}_+\right|^2\right) +\left|D_1\right|^2 + \left|D_3\right|^2 + \left|D_4\right|^2 +  \left|T_1\right|^2 - \left|T_4\right|^2  = 0\,.
	\label{ddf4}
\end{align}
Consequently, the solution \eqref{flata} has enough free parameters as to allow compliance with the D-flatness requirements. 

Let us now turn to the tree-level singlet field masses. A detailed calculation shows that any solution involving non-vanishing VEVs for all singlet fields not included in \eqref{flata}
renders $\Phi_{12}, \oPhi_{12}, \Phi_3, \Phi_5$ as well as four linear combinations of $\phi_3,\ophi_3,\phi_4,\ophi_4,\phi_+,\ophi_+,\phi_-,\ophi_-, \phi_{45}, \ophi_{45}$,  $\Phi_{23}$
superheavy. Focusing on mass-mixings involving the field $\phi_3$, we have only two relevant superpotential terms 
\begin{align}
w_{3m} \sim \Phi_5\left(\phi_3\ophi_4+\phi_4\ophi_3\right)
\label{w3m}
\end{align}
that result to a massless $\phi_0$ and a massive $\phi_m$ linear combinations
\begin{align}\label{phi0}
\phi_0 = \sin\omega\,\phi_3-\cos\omega\,\ophi_3\,,\\
\phi_m = \cos\omega\,\phi_3+\sin\omega\,\ophi_3\,,
\end{align}
 where $\tan\omega={\langle\phi_4\rangle}/{\langle\ophi_4\rangle}$. In view of this, we identify the inflaton field $y$ with the massless combination $\phi_0$.
Solving backwards, we find that the inflaton participates in interactions via the fields
\begin{align}
\phi_3 = \sin\omega\,\phi_0 + \dots\quad;\quad 
\ophi_3 = -\cos\omega\,\phi_0 + \dots\label{imix}
\end{align}
where the dots stand for the superheavy combination $\phi_m$.
%
%
%

Of key importance to our analysis are the higher order non-renormalisable (NR) corrections to the 
superpotential \eqref{wtree}. A generic non-renormalisable term at order $N>3$  is of the form
\begin{align}
g_s^{N-2} C_N\,\varphi_1 \varphi_2 \varphi_3 \dots \varphi_N 
\,,
\end{align}
in units $\sqrt{2\alpha'}=1$, where 
 $C_N$ is a numerical constant proportional to the correlator
\begin{align}
\langle \varphi_1^F \varphi_2^F \varphi_3^B\dots\varphi_N^B \rangle\,.
\end{align} 
Here, $\varphi_i^F/\varphi_i^B$ stand for the fermionic/bosonic vertex operators of the superfield $\varphi_i$. The constant $C_N$ can be computed explicitly and it turns out to be either zero or of order one \cite{Kalara:1990fb, Cleaver:1997nj},\cite{Cvetic:1998gv}.\footnote{Actually $C_3=\sqrt{2}$, while $C_N$ with $N\ge 4$ was found to be of order 10 for the particular examples it was computed.} The vanishing of $C_N$ is controlled by gauge invariance,  Ising correlators related to real fermionic internal coordinates, and additional selection rules ascribed to $N=2$ worldsheet superalgebra\cite{Rizos:1991bm}. As the number of candidate NR terms is growing fast with the order $N$,
a systematic application of all selection criteria is only possible via the use of a computer code.
In our analysis we have used a specialised computer program to calculate all non-vanishing NR contributions to the superpotential up to sixth order. 

To order $N=6$ the F-flatness conditions, after applying \eqref{flata} and choosing
\begin{align}
\Phi_2 = T_1^2 = T_3^2 = T_3{\cdot}T_4 =  D_3 = D_1^2 = D_4^2= 0\,,
\label{dddd}
\end{align}
read
\begin{align}
\left(\Phi_{12}\right) &: \phi_4^2+\phi_+\phi_- + 
\left\{\left(D_1{\cdot}D_5\right)^2
+ \left(D_4{\cdot}D_5\right)^2\right\} = 0\,,\label{fPhi12}\\
(\oPhi_{12}) &: \ophi_4^2 + \ophi_+ \ophi_- =0\,,\label{ofPhi12}\\
\left(\Phi_3\right) &: \phi_4\ophi_4 + \phi_{45} \ophi_{45} + \phi_+\ophi_+ + \phi_-\ophi_- = 0\,,\label{Phi3}\\
\left(\ophi_3\right) &: D_4{\cdot}D_5  +\left\{\left(D_1{\cdot}D_5\right)\left(T_1{\cdot}T_4\right)\right\}= 0\,,\label{fphi3}\\
\left(T_1\right) &:
\left\{\left(\ophi_4^2 + \ophi_+ \ophi_-\right) \Phi_{31} T_1\right\} = 0\,,\label{fhoa}\\
\left(D_1\right)  &: \left\{\left(\ophi_4^2 + \ophi_+ \ophi_-\right)\Phi_{31} D_1  \right\}= 0\,,\label{fhob}\\
\left(D_4\right) &:
\left\{\left(\ophi_4^2 + \ophi_+ \ophi_-\right)\Phi_{31} D_4 \right\} = 0\,,\label{fhoc}
\end{align}
where we have omitted for notational simplicity order one numerical coefficients for terms arising from NR interactions (terms in curly brackets).
It is possible to extend the tree-level solution \eqref{flata}, \eqref{etf}, \eqref{etl} as to include up to order $N=6$ NR corrections. 
To this end, we remark that $\xi$ defined in \eqref{xidef} is the only explicit parameter in the system of flatness constraints.
In addition, $\xi$ is of the order of $10^{-1}-10^{-2}$ in string units. Hence, we can consider solutions expressed as power series in the small parameter $\xi$, starting from a tree-level expression solving Eq.~\eqref{etl} at the leading order in this series expansion. In this regard, we can solve the $\left(\ophi_3\right)$ flatness \eqref{fphi3} 
assuming $D_4{\cdot}D_5$ is of the order $\xi^4$
\begin{align}
D_4{\cdot}D_5 = - \left\{\left(D_1{\cdot}D_5\right)\left(T_1{\cdot}T_4\right)\right\} \sim \xi^4\,.
\end{align}
Then, we can also solve Eq. \eqref{fPhi12}
perturbatively 
\begin{align}
\phi_4^2+\phi_+\phi_- = 
\left\{\left(D_1{\cdot}D_5\right)^2\right\} \sim \xi^4\,. 
\label{d1d5nr}
\end{align}
To solve Eqs. \eqref{fhoa}-\eqref{fhoc} together with \eqref{ofPhi12} we also choose
\begin{align}
\label{phibars}
\ophi_4^2 = - \ophi_+ \ophi_- \sim \xi^4\,,
\end{align}
which satisfies the above F-flatness conditions to the order considered here (up to $N=6$), at it effectively diminishes the associated contributions in Eqs. \eqref{fhoa}-\eqref{fhoc} down to $N=7$ order.

Having solved the flatness conditions, let us now consider higher order contributions to the singlet masses and mixings. 
More particularly, we focus first on bilinear couplings comprising the inflaton field $\phi_0$.
Using eq.~\eqref{phi0}, these correspond to interactions of the form $\left(\phi_3\varphi\right) \varphi_1\dots\varphi_{N-2}$ and
 $\left(\ophi_3\varphi\right) \varphi_1\dots\varphi_{N-2}$, where $\varphi\in\left\{\Phi_I,\Phi_a,\oPhi_a,\phi_i,\ophi_i\right\},I=1,2,4,a=31,23,i=1,2,3,4,45$ and $\varphi_\alpha$ stand for 
singlet or hidden sector field VEVs. Taking into account our F/D-flatness solution we find a {\it single} non-vanishing such coupling up to and including $N=6$ order NR terms in the superpotential
\begin{align}
\left[\ophi_3 \Phi_4\right] \left(D_1{\cdot}D_5\right) \left(T_1{\cdot}T_4\right)\,.
\end{align}
For completeness, we also check all bilinear couplings of the same type involving $\Phi_4$.  We find a single mixing term at $N=6$:
\begin{align}
\left[\Phi_4\Phi_{12}\right]\left(D_1{\cdot}D_5\right)^2\,.
\end{align}
This coupling does not play an important r\^ole in our analysis as it simply introduces a small mixing of $\Phi_4$ with $\Phi_{12}$ that became superheavy at tree-level.
Following the discussion in Section 3, we are led to identify $\Phi_4$ with the goldstino superfield. 

Next,  we compute all trilinear couplings induced by NR terms of the form 
$\left(\phi_3^2\Phi_4\right) \varphi_1\dots\varphi_{N-3}$,  $\left(\ophi_3^2\Phi_4\right) \varphi_1\dots\varphi_{N-3}$ and $\left(\phi_3\ophi_3\Phi_4\right) \varphi_1\dots\varphi_{N-3}$. 
It turns out that up to and including $N=7$ no such coupling exists. At $N=8$, we find a {\it unique} term
\begin{align}
\left[\ophi_3^2\Phi_4\right]\left(D_1{\cdot}D_4\right) \left(T_1{\cdot}T_4\right)\Phi_{31}\,.
\end{align}
Altogether, the inflaton related NR effective superpotential reads
\begin{align}\label{WI}
W_I &= g_s 
C_6 \left(g_s\sqrt{2\alpha'}\right)^{3}\,\ophi_3 \Phi_4 \langle{D_1}\rangle{\cdot}\langle{D_5}\rangle\, \langle{T_1}\rangle{\cdot}\langle{T_4}\rangle\\
&\hspace{0.8cm}+g_s 
C_8 \left(g_s\sqrt{2\alpha'}\right)^{5}\,\ophi_3^2\Phi_4\langle{D_1}\rangle{\cdot}\langle{D_4}\rangle\, \langle{T_1}\rangle{\cdot}\langle{T_4}\rangle\langle{\Phi_{31}}\rangle\,,\nonumber
\end{align}
where $C_6,C_8$ stand for the numerical values of the correlators associated to the $N=6$ and $N=8$ NR couplings, respectively and we restored the mass units. 

One might wonder whether higher order corrections could destabilise the inflation mechanism based on the above-mentioned superpotential $W_I$. Higher order corrections are of order $\phi^n$ which are suppressed when the fields are small compared to the Planck scale. This is trivially true for all fields except possibly for the inflaton $y$
during the inflation period, where corrections may be important. This is actually a general problem of large field inflation, such as the Starobinsky model. In our case corrections are calculable in string theory, and we have checked that the inflationary potential found above is not destabilised at the next two orders $N=9,10$.

In order to make an estimate of the resulting inflaton superpotential, we introduce a typical VEV $\langle{\phi}\rangle$ as dummy variable and rewrite \eqref{WI}, using also \eqref{imix}, as:
\begin{align}
\label{WI2}
W_I= \zeta^4 \Phi_4\left(\frac{\gamma}{g_s\sqrt{2\alpha'}}\phi_0
+  \delta \zeta \phi_0^2 \right)\,,
\end{align}
where $\gamma, \delta$ and $\zeta$ are dimensionless parameters given by:
\begin{align}
\gamma &=  -g_s 
C_6 \frac{\langle{D_1}\rangle{\cdot}\langle{D_5}\rangle
\langle{T_1}\rangle{\cdot}\langle{T_4}\rangle}{\langle\phi\rangle^4}\cos\omega\,,\\
\delta &= \hphantom{+}g_s 
C_8 \frac{\langle{D_1}\rangle{\cdot}\langle{D_4}\rangle \langle{T_1}\rangle{\cdot}\langle{T_4}\rangle
\langle\Phi_{31}\rangle}{\langle\phi\rangle^5}\cos^2\omega\,, \label{eeqq}\\
\zeta &= {\langle\phi\rangle g_s\sqrt{2\alpha'}}\,.
\end{align}
A typical VEV $\langle{\phi}\rangle$ satisfying the D-term conditions \eqref{df1}-\eqref{xidef} is of order $\xi$, leading to an estimate:
\begin{align}
\zeta \sim {\xi g_s\sqrt{2\alpha'}} = \frac{g_s}{2\pi}\simeq 0.08\,,
\end{align}
where we used $g_s=g/\sqrt{2}$, with $g$ the GUT gauge coupling, corresponding to the standard convention 1/2 for the trace of the square of gauge group  generators in the fundamental representation.

At this point, one can easily reconstruct the Starobinsky-like inflationary potential of Section 3 from the superpotential \eqref{WI2}, by renaming $\phi_0=y, \Phi_4=z$:
\begin{align}
W_I= 
M_I z \left(y - \lambda y^2\right) \,,
\end{align}
where 
\be
\label{MI}
M_I=\zeta^4 {\gamma\over g_s}{1\over\sqrt{2\alpha'}}\simeq \zeta^4 C_6\cos\omega{1\over\sqrt{2\alpha'}}
\ee
is the inflation scale and 
\be
\lambda=-g_s\zeta{\delta\over\gamma}\sqrt{2\alpha'}M_P\simeq g_s\zeta {C_8\over C_6}
{\langle{D_1}\rangle{\cdot}\langle{D_4}\rangle\over \langle{D_1}\rangle{\cdot}\langle{D_5}\rangle}
\sqrt{2\alpha'}M_P\,.
\ee
Then, for $\gamma\sim \zeta\delta$, obtained for instance by choosing
\begin{align}
\omega\sim0\ ,\ \langle{D_1}\rangle{\cdot}\langle{D_5}\rangle\sim \zeta\langle\phi\rangle^2\ ,\ 
\langle{D_1}\rangle{\cdot}\langle{D_4}\rangle\sim \langle\phi\rangle^2\ ,\  \langle{T_1}\rangle{\cdot}\langle{T_4}\rangle\sim \langle\phi\rangle^2\ ,\ \langle\Phi_{31}\rangle\sim\langle\phi\rangle\,,
\end{align}
and using the relation between the reduced Planck mass $M_P=1/\sqrt{8\pi G}$ and the heterotic string scale $M_s\equiv 1/\sqrt{2\alpha'}$
\be\label{MP}
M_P^2= 
{2\over g_s^2\alpha'}\quad\Rightarrow\quad {1\over\sqrt{2\alpha'}}\simeq 5\times 10^{17}\,{\rm GeV}\,,
\ee
we obtain $M_I\simeq\zeta^5 C_6M_s\sim 10^{13}$ GeV and $\lambda\simeq g_sM_P/M_s$ an order one tunable parameter. It is curious to notice that by tuning $\lambda$ to unity, one can produce in the string spectrum a scalar that has exactly the properties of an $R^2$ term in the effective action.

Thus,
the effective superpotential $W_I$ is in full agreement with \eqref{Wlambda} 
for a range of VEVs of the singlet and hidden sector fields in the theory compatible with F/D-flatness constraints. Putting everything together we conclude that 
the Starobinsky-like inflationary scenario is fully realisable in the string derived ``revamped" flipped  ${SU(5)}\times{U(1)}$  model.

Note that the  $\lambda_6$ term  $F_4 {\bar F}_5 \phi_3$ in eq.~\eqref{wtree} 
metamorphoses, using eq.~\eqref{imix}, to $\sin\omega F_4{\bar F}_5\phi_0$, involving the inflaton field $\phi_0$ thus the $\lambda_6$ coupling becomes $\sin\omega/\sqrt{2}$. This is a highly desirable result, because as has been 
found in Refs.~\cite{Ellis:2020lnc, Ellis:2018moe} $\lambda_6$ should be $\simlt 10^{-3}$  in order to provide appropriate reheating, baryogenesis and light neutrino masses. Thus $\langle\phi_4\rangle/\langle{\bar\phi}_4\rangle$ should be $<<1\,$, compatible with the F- and D-flat directions imposed constraints on the gauge singlets VEVs. Indeed, the following solution to the flatness conditions satisfy this requirement:
\begin{align}
	\ophi_4\sim\xi^2\ ,\ \phi_-\sim\xi^3\ ,\ \ophi_+\sim\xi^3\ ,\ 
\end{align}
and 
$\phi_4/\ophi_4\sim 10^{-3}$.
This is compatible with \eqref{d1d5nr} (due to the presence of the NR term) and \eqref{phibars}, provided $\ophi_{45}\sim\xi^3$ in order to satisfy \eqref{etl}.
After applying \eqref{flata}, \eqref{dddd} and \eqref{fPhi12}-\eqref{fhoc} the D-flatness equations reduce to:
\begin{align}
	&\left|\phi_{45}\right|^2  -\frac{\left|T_3\right|^2}{2} \sim 
	\xi^2\,, \label{adf1}\\
	&\left|\ophi_-\right|^2 +  \left|{\phi}_+\right|^2 -\frac{\left|T_3\right|^2}{2} \sim \xi^2\,,\label{adf2}\\
	&\left|\Phi_{31}\right|^2 
	-\frac{1}{2}\left(\left|D_1\right|^2 + \left|D_4\right|^2 - \left|D_5\right|^2\right)
	-\frac{1}{2} \left(\left|T_1\right|^2  + \left|T_3\right|^2 - \left|T_4\right|^2 \right)
	= 3\xi^2\,,\label{addf3}\\
	&
	2 \left|\phi_+\right|^2
	+\left|D_1\right|^2 + \left|D_4\right|^2 +  \left|T_1\right|^2 - \left|T_4\right|^2  \sim 0\,,
	\label{addf4}
\end{align}
which can be easily satisfied.

\section{Conclusions and outook}

In summary, in this work, we have investigated the cosmology of a string derived flipped $SU(5)\times U(1)$ model in the free-fermionic framework of 4d heterotic superstrings. We have shown that a successful Starobinsky-type inflation can be accommodated with the inflaton being the superpartner of a state that mixes with a right-handed neutrino (linear combination) and an appropriate superpotential involving also the goldstino superfield. The effective field theory is a $N=1$ no-scale supergravity with an exactly calculable (to all orders in $\alpha'$) K\"ahler potential and a superpotential induced perturbatively around the fermionic point via a set of VEVs for $SU(5)\times U(1)$ singlet fields, driven by an anomalous $U(1)_A$ gauge symmetry generally present in chiral heterotic string constructions 


An important point is that during inflation, the GUT symmetry group $SU(5)\times U(1)$ remains unbroken; its breaking occurs via a first order phase transition at lower energies, after the end of inflation. This is a desirable feature, since the present analysis of the inflationary period does not a priori affect the $SU(5)\times U(1)$ breaking vacuum and the resulting particle phenomenology that has been studied extensively in the past. Another related feature is that soft supersymmetry breaking terms do not play any r\^ole  during inflation, with the inflaton rolling down to a supersymmetric minimum before the phase transition to the low energy vacuum occurs.

In this paper, we performed a detailed analysis of the non-renormalisable terms in the superpotential, up to dimension-eight, and identified a set of $SU(5)\times U(1)$ singlet VEVs consistent with the F- and D-flatness conditions, that generate the requested inflationary superpotential for the inflaton and goldstino superfields; the latter were appropriately chosen among the gauge singlets of the model with the desired K\"ahler potential, so that the resulting scalar potential leads to a Starobinsky-type inflation consistent with the CMB observations.

Several open questions remain to be studied. In particular, a stabilisation mechanism for the dilaton is desirable in heterotic string models using either non-perturbative effects and/or magnetic and 3-form fluxes in analogy with type II models. 
Obviously, our result is a proof of concept showing the existence of a general class of models sharing similar properties. On the other hand, it is astonishing that the explicit revamped flipped $SU(5)$ construction, built 30 years ago, accommodates remarkably well a low energy phenomenology both in particle physics and cosmology. A detailed phenomenological analysis combining all aspects is currently under investigation.

\section*{Acknowledgements}

J.R. would like to thank LPTHE for hospitality and financial support.
Work supported in part by a CNRS-PICS grant 07964. The work of DVN was supported in part by the DOE grant DE-FG02-13ER42020 at Texas A\&M University and in part by the Alexander S. Onassis Public Benefit Foundation. 
We would like to thank Natsumi Nagata and especially Keith Olive for enlightening discussions.

\newpage
\begin{appendices}
\section{String construction and spectrum of the flipped $SU(5)\times{U(1)}$ model}
The revamped flipped $SU(5)\times{U(1)}$ string model is constructed using the free fermionic formulation of the heterotic
superstring. In this framework a string model is defined by a set of basis vectors $B=\left\{\beta_1,\beta_2,\dots,\beta_n\right\}$ and a set of phases $C_{ij}=c\left[\beta_i\atop\beta_j\right]\,,i,j=1,\dots,n$. The former is associated with the 
transformation properties of the world-sheet fermions while the later is related to generalised GSO projections.
The revamped flipped model is generated using 8 basis vectors, namely
\begin{align}
\beta_1 =S & = \left\{\psi^\mu,\chi^{1,\dots,6}\right\}\,,\nonumber\\
\beta_2 =b_1 & = \left\{\psi^\mu,\chi^{12},y^{3,\dots,6};\overline{y}^{3,\dots,6},\overline{\psi}^{1,\dots,5},\overline{\eta}^1\right\}\,,\nonumber\\
\beta_3 =b_2 & = \left\{\psi^\mu,\chi^{34},y^{12},\omega^{56};\overline{y}^{12},\overline{\omega}^{56},\overline{\psi}^{1,\dots,5}\overline{\eta}^2\right\}\,,\nonumber\\
\beta_4 =b_3 & = \left\{\psi^\mu,\chi^{56},\omega^{1,\dots,4};\overline{\omega}^{1,\dots,4},\overline{\psi}^{1,\dots,5},\overline{\eta}^3\right\}\,,\\
\beta_5 =b_4 & = \left\{\psi^\mu,\chi^{12},y^{36},\omega^{45};\overline{y}^{36},\overline{\omega}^{45},\overline{\psi}^{1,\dots,5},\overline{\eta}^1\right\}\,,\nonumber\\
\beta_6 =b_5 & = \left\{\psi^\mu,\chi^{34},y^{26},\omega^{15};\overline{y}^{26},\overline{\omega}^{15},\overline{\psi}^{1,\dots,5},\overline{\eta}^2\right\}\,,\nonumber\\
\beta_7 =\zeta & = \left\{\overline{\phi}^{1,\dots,8}\right\}\,,\nonumber\\
\beta_8 =\alpha & = \Bigl\{y^{46},\omega^{46};\overline{y}^{46},\overline{\omega}^{2346},
\underbrace{\overline{\psi}^{1,\dots,5},\overline{\eta}^{1,2,3},\overline{\phi}^{1,\dots,4}}_{\frac{1}{2},\dots,\frac{1}{2}},\overline{\phi}^{56}\Bigr\}\,,\nonumber
\end{align}
where included fermions are periodic unless indicated otherwise (i.e. $\frac{1}{2}$ stands for fermions twisted by $-i$). The associated GGSO phase matrix elements are $C_{ij} = c\left[\beta_i\atop\beta_j\right]=\exp(i\pi \tilde{c}_{ij})$, where
\begin{align}
\tilde{c} = 
\bordermatrix{~&S&b_1&b_2&b_3&b_4&b_5&\zeta&\alpha\cr
S&0&0&0&0&0&0&1&1\cr
b_1&1&1&1&1&1&1&1&-1/2\cr
b_2&1&1&1&1&1&1&1&-1/2\cr
b_3&1&1&1&1&1&1&1&1\cr
b_4&1&1&1&1&1&1&1&-1/2\cr
b_5&1&1&1&1&1&1&1&+1/2\cr
\zeta&1&1&1&1&1&1&1&1\cr
\alpha&1&1&1&1&1&0&1&-1/2\cr
}\,.
\end{align}
The above set of basis vectors and GGSO phases define a $N=1$ supersymmetric string model exhibiting 
$SU(5)\times{U(1)}\times{U(1)}^4\times{SU(4)}\times{SO(10)}$ gauge symmetry. We will consider the
flipped $SU(5)\times{U(1)}$  group factor as the ``observable" gauge group and the ${SU(4)}\times{SO(10)}$ group factor as the 
``hidden" gauge group. The four additional abelian factors ${U(1)}^4=\prod_{i=1}^4{U(1)}_i$ corresponding to the world-sheet currents
$\oeta^1\oeta^{1\ast},\oeta^2\oeta^{1\ast},\oeta^3\oeta^{3\ast}, \overline{\omega}^2\overline{\omega}^3$ 
appear to be anomalous; ${\rm Tr}{U(1)}_1=-36, {\rm Tr}{U(1)}_2=-12, {\rm Tr}{U(1)}_3=-24, {\rm Tr}{U(1)}_4=-12$. Redefining appropriately we obtain three  $U(1)$ combinations free of gauge and mixed gravitational anomalies
\begin{align}
{U(1)}'_1 &= {U(1)}_3+2U(1)_4\,,\\
{U(1)}'_2 &= {U(1)}_1-3U(1)_2\,,\\
{U(1)}'_3 &= 3{U(1)}_1+{U(1)}_2+4{U(1)}_3-2{U(1)}_4\,,
\end{align}
and one anomalous abelian group factor
\begin{align}
{U(1)}_A = -3 {U(1)}_1 - {U(1)}_2+2{U(1)}_3-{U(1)}_4\ \,,\ {\rm Tr}{U(1)}_A = 180\,.
\end{align}

The massless matter spectrum comprises:
(i) ``Observable" sector matter particles, that is, states charged solely under the $SU(5)\times{U(1)}\times{U(1)}^4$ group factor, listed in Table \ref{tobs}. These include three chiral fermion families as well as a family/anti-family pair, accommodated in $SO(10)$ spinorials/anti-spinorials, arising from the sectors $b_1,b_2,b_3,b_4,b_5$, and three fiveplet/anti-fiveplet pairs, accommodated in $SO(10)$ vectorials, from the sectors $S, S+b_4+b_5$. Bosonic partners come from the sectors $S+b_i, i=1,\dots,4$ and $0, b_4+b_5$ respectively.
(ii) ``Hidden" sector matter particles, that is, states exclusively charged under ${SU(4)}\times{SO(10)}\times{U(1)}^4$ listed in Table \ref{this}. These arise from the sectors
\footnote{Here we employ a compact notation to denote several sectors contributing to the same field multiplet, for example  $(S)+ b_i +2\alpha(+\zeta)$ stands for four sectors: $b_i+2\alpha, b_i+2\alpha+\zeta, S+b_i+2\alpha,  S+b_i+2\alpha+\zeta$.}
 $(S)+b_i+2\alpha (+\zeta), i=1,\dots,4$.
(iii) Exotic fractionally charged states coming from the sectors $(S)+b_1\pm\alpha (+\zeta)$, $(S)+b_1+b_4+b_5\pm\alpha (+\zeta)$,
$(S)+b_2+b_3+b_5\pm\alpha (+\zeta)$, $(S)+b_1+b_2+b_4\pm\alpha (+\zeta)$. $(S)+b_2+b_4\pm\alpha (+\zeta)$, $(S)+b_4\pm\alpha (+\zeta)$. These are listed in Table \ref{texo}.

\begin{table}[!ht]
\centering
{\footnotesize
\begin{tabular}{|l|c|c||c|c|c|c||c|c|}
\hline
&$SU(5)$&$U(1)$&${U(1)}_1$&${U(1)}_2$&${U(1)}_3$&${U(1)}_4$&$SU(4)$&$SO(10)$\\
\hline
\multicolumn{9}{|l|}{$S$}\\
\hline
$h_1$&
${\mathbf{5}}$&$-1$&$+1$&$0$&$0$&$0$&${\mathbf{1}}$&$\mathbf{1}$\\
\hline
$\bar{h}_1$&${\overline{\mathbf{5}}}$&$+1$&$-1$&$0$&$0$&$0$&${\mathbf{1}}$&$\mathbf{1}$\\
\hline
$h_2$&
${\mathbf{5}}$&$-1$&$0$&$+1$&$0$&$0$&${\mathbf{1}}$&$\mathbf{1}$\\
\hline
$\bar{h}_2$&
${\overline{\mathbf{5}}}$&$+1$&$0$&$-1$&$0$&$0$&${\mathbf{1}}$&$\mathbf{1}$\\
\hline
$h_3$&
${\mathbf{5}}$&$-1$&$0$&$0$&$+1$&$0$&${\mathbf{1}}$&$\mathbf{1}$\\
\hline
$\bar{h}_3$&
${\overline{\mathbf{5}}}$&$+1$&$0$&$0$&$-1$&$0$&${\mathbf{1}}$&$\mathbf{1}$\\
\hline
$\Phi_{12}$&
${\mathbf{1}}$&$0$&$-1$&$+1$&$0$&$0$&${\mathbf{1}}$&$\mathbf{1}$\\
\hline
$\overline{\Phi}_{12}$&
${\mathbf{1}}$&$0$&$+1$&$-1$&$0$&$0$&${\mathbf{1}}$&$\mathbf{1}$\\
\hline
${\Phi}_{31}$&
${\mathbf{1}}$&$0$&$+1$&$0$&$-1$&$0$&${\mathbf{1}}$&$\mathbf{1}$\\
\hline
$\overline{\Phi}_{31}$&
${\mathbf{1}}$&$0$&$-1$&$0$&$+1$&$0$&${\mathbf{1}}$&$\mathbf{1}$\\
\hline
$\Phi_{23}$&
${\mathbf{1}}$&$0$&$0$&$-1$&$+1$&$0$&${\mathbf{1}}$&$\mathbf{1}$\\
\hline
$\overline{\Phi}_{23}$&
${\mathbf{1}}$&$0$&$0$&$+1$&$-1$&$0$&${\mathbf{1}}$&$\mathbf{1}$\\
\hline
$\Phi_I\,,I=1,\dots,5$&
${\mathbf{1}}$&$0$&$0$&$0$&$0$&$0$&${\mathbf{1}}$&$\mathbf{1}$\\
\hline
\multicolumn{9}{|l|}{$b_1$}\\
\hline
$F_1$&
${\mathbf{10}}$&$+\mfrac{1}{2}$&$-\mfrac{1}{2}$&$0$&$0$&$0$&${\mathbf{1}}$&$\mathbf{1}$\\
\hline
$\overline{f}_1$&
${\overline{\mathbf{5}}}$&$-\mfrac{3}{2}$&$-\mfrac{1}{2}$&$0$&$0$&$0$&${\mathbf{1}}$&$\mathbf{1}$\\
\hline
${\ell}^c_1$&
${\mathbf{1}}$&$+\mfrac{5}{2}$&$-\mfrac{1}{2}$&$0$&$0$&$0$&${\mathbf{1}}$&$\mathbf{1}$\\
\hline
\multicolumn{9}{|l|}{$b_2$}\\
\hline
$F_2$&
${\mathbf{10}}$&$+\mfrac{1}{2}$&$0$&$-\mfrac{1}{2}$&$0$&$0$&${\mathbf{1}}$&$\mathbf{1}$\\
\hline
$\overline{f}_2$&
${\overline{\mathbf{5}}}$&$-\mfrac{3}{2}$&$0$&$-\mfrac{1}{2}$&$0$&$0$&${\mathbf{1}}$&$\mathbf{1}$\\
\hline
${\ell}^c_2$&
${\mathbf{1}}$&$+\mfrac{5}{2}$&$0$&$-\mfrac{1}{2}$&$0$&$0$&${\mathbf{1}}$&$\mathbf{1}$\\
\hline
\multicolumn{9}{|l|}{$b_3$}\\
\hline
$F_3$&
${\mathbf{10}}$&$+\mfrac{1}{2}$&$0$&$0$&$+\mfrac{1}{2}$&$-\mfrac{1}{2}$&${\mathbf{1}}$&$\mathbf{1}$\\
\hline
$\overline{f}_3$&
${\overline{\mathbf{5}}}$&$-\mfrac{3}{2}$&$0$&$0$&$+\mfrac{1}{2}$&$+\mfrac{1}{2}$&${\mathbf{1}}$&$\mathbf{1}$\\
\hline
${\ell}^c_3$&
${\mathbf{1}}$&$+\mfrac{5}{2}$&$0$&$0$&$+\mfrac{1}{2}$&$+\mfrac{1}{2}$&${\mathbf{1}}$&$\mathbf{1}$\\
\hline
\multicolumn{9}{|l|}{$b_4$}\\
\hline
$F_4$&
${\mathbf{10}}$&$+\mfrac{1}{2}$&$-\mfrac{1}{2}$&$0$&$0$&$0$&${\mathbf{1}}$&$\mathbf{1}$\\
\hline
${f}_4$&
${{\mathbf{5}}}$&$+\mfrac{3}{2}$&$+\mfrac{1}{2}$&$0$&$0$&$0$&${\mathbf{1}}$&$\mathbf{1}$\\
\hline
$\overline{\ell}^c_4$&
${\mathbf{1}}$&$-\mfrac{5}{2}$&$+\mfrac{1}{2}$&$0$&$0$&$0$&${\mathbf{1}}$&$\mathbf{1}$\\
\hline
\multicolumn{9}{|l|}{$b_5$}\\
\hline
$\oF_5$&
$\overline{\mathbf{10}}$&$-\mfrac{1}{2}$&$0$&$+\mfrac{1}{2}$&$0$&$0$&${\mathbf{1}}$&$\mathbf{1}$\\
\hline
$\overline{f}_5$&
${{\mathbf{5}}}$&$-\mfrac{3}{2}$&$0$&$-\mfrac{1}{2}$&$0$&$0$&${\mathbf{1}}$&$\mathbf{1}$\\
\hline
${\ell}^c_5$&
${\mathbf{1}}$&$+\mfrac{5}{2}$&$0$&$-\mfrac{1}{2}$&$0$&$0$&${\mathbf{1}}$&$\mathbf{1}$\\
\hline
\multicolumn{9}{|l|}{$S+b_4+b_5$}\\
\hline
$h_{45}$&
${\mathbf{5}}$&$-1$&$-\mfrac{1}{2}$&$-\mfrac{1}{2}$&$0$&$0$&${\mathbf{1}}$&$\mathbf{1}$\\
\hline
$\overline{h}_{45}$&
$\overline{\mathbf{5}}$&$+1$&$+\mfrac{1}{2}$&$+\mfrac{1}{2}$&$0$&$0$&${\mathbf{1}}$&$\mathbf{1}$\\
\hline
${\phi}_{45}$&${{\mathbf{1}}}$&$0$&$+\mfrac{1}{2}$&$+\mfrac{1}{2}$&$+1$&$0$&${\mathbf{1}}$&$\mathbf{1}$\\
\hline
$\overline{\phi}_{45}$&${{\mathbf{1}}}$&$0$&$-\mfrac{1}{2}$&$-\mfrac{1}{2}$&$-1$&$0$&${\mathbf{1}}$&$\mathbf{1}$\\
\hline
${\phi}_{+}$&${{\mathbf{1}}}$&$0$&$+\mfrac{1}{2}$&$-\mfrac{1}{2}$&$0$&$+1$&${\mathbf{1}}$&$\mathbf{1}$\\
\hline
$\overline{\phi}_{+}$&${{\mathbf{1}}}$&$0$&$-\mfrac{1}{2}$&$+\mfrac{1}{2}$&$0$&$-1$&${\mathbf{1}}$&$\mathbf{1}$\\
\hline
${\phi}_{-}$&${{\mathbf{1}}}$&$0$&$+\mfrac{1}{2}$&$-\mfrac{1}{2}$&$0$&$-1$&${\mathbf{1}}$&$\mathbf{1}$\\
\hline
$\overline{\phi}_{-}$&${{\mathbf{1}}}$&$0$&$-\mfrac{1}{2}$&$+\mfrac{1}{2}$&$0$&$+1$&${\mathbf{1}}$&$\mathbf{1}$\\
\hline
${\phi}_{i}\,,i=1,\dots,4$&${{\mathbf{1}}}$&$0$&$+\mfrac{1}{2}$&$-\mfrac{1}{2}$&$0$&$0$&${\mathbf{1}}$&$\mathbf{1}$\\
\hline
$\overline{\phi}_{i}\,, i=1,\dots,4$&${{\mathbf{1}}}$&$0$&$-\mfrac{1}{2}$&$+\mfrac{1}{2}$&$0$&$0$&${\mathbf{1}}$&$\mathbf{1}$\\
\hline
\end{tabular}
}
\caption{``Observable" sector massless matter states and their $SU(5)\times{U(1)}\times{U(1)}^4\times{SU(4)}\times{SO(10)}$ quantum numbers.}
\label{tobs}
\end{table}

\begin{table}[!ht]
\centering
{\footnotesize
\begin{tabular}{|l|c|c||c|c|c|c||c|c|}
\hline
&$SU(5)$&$U(1)$&${U(1)}_1$&${U(1)}_2$&${U(1)}_3$&${U(1)}_4$&$SU(4)$&$SO(10)$\\
\hline
\multicolumn{9}{|l|}{$b_1+2\alpha\,\left(+\zeta\right)$}\\
\hline
$D_1$&
${\mathbf{1}}$&$0$&$0$&$-\mfrac{1}{2}$&$+\mfrac{1}{2}$&$0$&${\mathbf{6}}$&$\mathbf{1}$\\
\hline
$T_1$&
${\mathbf{1}}$&$0$&$0$&$-\mfrac{1}{2}$&$+\mfrac{1}{2}$&$0$&${\mathbf{1}}$&$\mathbf{10}$\\
\hline
\multicolumn{9}{|l|}{$b_2+2\alpha\,\left(+\zeta\right)$}\\
\hline
$D_2$&
${\mathbf{1}}$&$0$&$-\mfrac{1}{2}$&$0$&$+\mfrac{1}{2}$&$0$&${\mathbf{6}}$&$\mathbf{1}$\\
\hline
$T_2$&
${\mathbf{1}}$&$0$&$-\mfrac{1}{2}$&$0$&$+\mfrac{1}{2}$&$0$&${\mathbf{1}}$&$\mathbf{10}$\\
\hline
\multicolumn{9}{|l|}{$b_3+2\alpha\,\left(+\zeta\right)$}\\
\hline
$D_3$&
${\mathbf{1}}$&$0$&$-\mfrac{1}{2}$&$-\mfrac{1}{2}$&$0$&$+\mfrac{1}{2}$&${\mathbf{6}}$&$\mathbf{1}$\\
\hline
$T_3$&
${\mathbf{1}}$&$0$&$-\mfrac{1}{2}$&$-\mfrac{1}{2}$&$0$&$-\mfrac{1}{2}$&${\mathbf{1}}$&$\mathbf{10}$\\
\hline
\multicolumn{9}{|l|}{$b_4+2\alpha\,\left(+\zeta\right)$}\\
\hline
$D_4$&
${\mathbf{1}}$&$0$&$0$&$-\mfrac{1}{2}$&$+\mfrac{1}{2}$&$0$&${\mathbf{6}}$&$\mathbf{1}$\\
\hline
$T_4$&
${\mathbf{1}}$&$0$&$0$&$+\mfrac{1}{2}$&$-\mfrac{1}{2}$&$0$&${\mathbf{1}}$&$\mathbf{10}$\\
\hline
\multicolumn{9}{|l|}{$b_5+2\alpha\,\left(+\zeta\right)$}\\
\hline
$D_5$&
${\mathbf{1}}$&$0$&$+\mfrac{1}{2}$&$0$&$-\mfrac{1}{2}$&$0$&${\mathbf{6}}$&$\mathbf{1}$\\
\hline
$T_5$&
${\mathbf{1}}$&$0$&$-\mfrac{1}{2}$&$0$&$+\mfrac{1}{2}$&$0$&${\mathbf{1}}$&$\mathbf{10}$\\
\hline
\end{tabular}
}
\caption{``Hidden sector" massless matter states and their $SU(5)\times{U(1)}\times{U(1)}^4\times{SU(4)}\times{SO(10)}$ quantum numbers.}
\label{this}
\end{table}

\begin{table}[!ht]
\centering
{\footnotesize
\begin{tabular}{|l|c|c||c|c|c|c||c|c|}
\hline
&$SU(5)$&$U(1)$&${U(1)}_1$&${U(1)}_2$&${U(1)}_3$&${U(1)}_4$&$SU(4)$&$SO(10)$\\
\hline
\multicolumn{9}{|l|}{$b_1\pm\alpha\,\left(+\zeta\right)$}\\
\hline
$\overline{X}_1$&
${\mathbf{1}}$&$-\mfrac{5}{4}$&$-\mfrac{1}{4}$&$+\mfrac{1}{4}$&$+\mfrac{1}{4}$&$+\mfrac{1}{2}$&$\overline{\mathbf{4}}$&$\mathbf{1}$\\
\hline
$\overline{X}_2$&
${\mathbf{1}}$&$-\mfrac{5}{4}$&$-\mfrac{1}{4}$&$+\mfrac{1}{4}$&$+\mfrac{1}{4}$&$-\mfrac{1}{2}$&$\overline{\mathbf{4}}$&$\mathbf{1}$\\
\hline
\multicolumn{9}{|l|}{$b_1+b_4+b_5\pm\alpha\,\left(+\zeta\right)$}\\
\hline
${Y}_1$&
${\mathbf{1}}$&$+\mfrac{5}{4}$&$-\mfrac{1}{4}$&$+\mfrac{1}{4}$&$-\mfrac{1}{4}$&$+\mfrac{1}{2}$&${\mathbf{4}}$&$\mathbf{1}$\\
\hline
${Y}_2$&
${\mathbf{1}}$&$+\mfrac{5}{4}$&$-\mfrac{1}{4}$&$+\mfrac{1}{4}$&$-\mfrac{1}{4}$&$-\mfrac{1}{2}$&${\mathbf{4}}$&$\mathbf{1}$\\
\hline
\multicolumn{9}{|l|}{$b_2+b_3+b_5\pm\alpha\,\left(+\zeta\right)$}\\
\hline
${Z}_2$&
${\mathbf{1}}$&$+\mfrac{5}{4}$&$-\mfrac{1}{4}$&$+\mfrac{3}{4}$&$+\mfrac{1}{4}$&$0$&${\mathbf{4}}$&$\mathbf{1}$\\
\hline
$\overline{Z}_2$&
${\mathbf{1}}$&$-\mfrac{5}{4}$&$-\mfrac{3}{4}$&$+\mfrac{1}{4}$&$-\mfrac{1}{4}$&$0$&$\overline{\mathbf{4}}$&$\mathbf{1}$\\
\hline
\multicolumn{9}{|l|}{$b_1+b_2+b_4\pm\alpha\,\left(+\zeta\right)$}\\
\hline
${Y}_2'$&
${\mathbf{1}}$&$+\mfrac{5}{4}$&$-\mfrac{1}{4}$&$+\mfrac{1}{4}$&$-\mfrac{1}{4}$&$-\mfrac{1}{2}$&${\mathbf{4}}$&$\mathbf{1}$\\
\hline
$\overline{Y}_1$&
${\mathbf{1}}$&$-\mfrac{5}{4}$&$+\mfrac{1}{4}$&$-\mfrac{1}{4}$&$+\mfrac{1}{4}$&$-\mfrac{1}{2}$&$\overline{\mathbf{4}}$&$\mathbf{1}$\\
\hline
\multicolumn{9}{|l|}{$S+b_2+b_4\pm\alpha\,\left(+\zeta\right)$}\\
\hline
${Z}_1$&
${\mathbf{1}}$&$-\mfrac{5}{4}$&$+\mfrac{1}{4}$&$+\mfrac{1}{4}$&$-\mfrac{1}{4}$&$+\mfrac{1}{2}$&${\mathbf{4}}$&$\mathbf{1}$\\
\hline
$\overline{Z}_1$&
${\mathbf{1}}$&$+\mfrac{5}{4}$&$-\mfrac{1}{4}$&$-\mfrac{1}{4}$&$+\mfrac{1}{4}$&$-\mfrac{1}{2}$&$\overline{\mathbf{4}}$&$\mathbf{1}$\\
\hline
\multicolumn{9}{|l|}{$b_4\pm\alpha\,\left(+\zeta\right)$}\\
\hline
${X}_1$&
${\mathbf{1}}$&$+\mfrac{5}{4}$&$+\mfrac{1}{4}$&$-\mfrac{1}{4}$&$-\mfrac{1}{4}$&$-\mfrac{1}{2}$&${\mathbf{4}}$&$\mathbf{1}$\\
\hline
$\overline{X}_2'$&
${\mathbf{1}}$&$-\mfrac{5}{4}$&$-\mfrac{1}{4}$&$+\mfrac{1}{4}$&$+\mfrac{1}{4}$&$-\mfrac{1}{2}$&$\overline{\mathbf{4}}$&$\mathbf{1}$\\
\hline
\end{tabular}
}
\caption{Exotic fractionally charged massless matter states and their $SU(5)\times{U(1)}\times{U(1)}^4\times{SU(4)}\times{SO(10)}$ quantum numbers.}
\label{texo}
\end{table}

\clearpage

\section{Calculation of the potential in no-scale supergravity}
Using the following K\"ahler potential and superpotential
\begin{align}
K&=-2\ln\left(1-\frac{1}{2}|z|^2\right)-2\ln\left(1-\left|y\right|^2\right)\,,\\
W&= \mu z \left(y-\lambda y^2\right)\,,
\end{align}
we get
\begin{align}
K_z&=\frac{\bar{z}}{1-\frac{\left|z\right|^2}{2}}\,,\\
K_y&=\frac{2\bar{y}}{1-\left|y\right|^2}\,,\\
K_{z\bar{z}}&=\frac{1}{\left(1-\frac{\left|z\right|^2}{2}\right)^2}\,,\\
K_{y\bar{y}}&=\frac{2}{\left(1-\left|y\right|^2\right)^2}\,,
\end{align}
and
\begin{align}
D_z W &= \mu y \frac{\left(2+\left|z\right|^2\right)\left(1-\lambda y\right)}{2-\left|z\right|^2}\,,\\
D_y W &= \mu z \frac{1-2 \lambda y + \left|y\right|^2}{1-\left|y\right|^2}\,.
\end{align}
The potential reads
\begin{align}
V &= e^{K} \left(\left|D_i W\right|^2-3\left|W\right|^2\right)
= \frac{\mu^2 P(y,z)}{\left(1-\left|y\right|^2\right)^2\left(2-\left|z\right|^2\right)^2}\,,
\end{align}
where we omited the K\"ahler metric $G_{i\bar j}\equiv K_{i\bar j}$ for notational simplicity in the contraction of indices, 
$|D_iW|^2\equiv G^{i\bar j}(D_iW)(D_{\bar j}{\overline W})$, and
\begin{align}
P&\!\!=\!\!\left\{\strut4|y|^2\left(1\!-\!2\lambda|y|\cos\theta\!+\!\lambda^2|y|^2\right)
+2\left(1\!-\!\left|y\right|^2\right)\left(1\!-\!|y|^2\!+\!4\lambda^2\left|y\right|^2\!-\! 4 \lambda \left|y\right| \cos\theta\right)\left|z\right|^2\right.\nonumber\\
&\left.\ \ \ + \left|y\right|^2 \left(1-2\lambda|y|\cos\theta+\lambda^2|y|^2\right)\left|z\right|^4
\right\}\,.
\end{align}
Minimizing with respect to $\theta$ we obtain potential minima for
\begin{align}
\cos\theta=\text{sgn}(\lambda)\,.
\end{align}
In the region $\left|y\right|\sim1$  the potential diverges unless the numerator is proportional to $\left(1-\left|y\right|\right)$. The remainder of the division of the numerator $P(y,z)$ over $\left(1-\left|y\right|\right)$ is 
\begin{align}
\mu^2 (4+\left|z\right|^4)\left|\left(1-2\lambda\cos\theta+\lambda^2\right)\right|_{\cos\theta=\text{sgn}
\left(\lambda\right)}=\mu^2 (4+\left|z\right|^4)1\left(1-|\lambda|\right)^2
\,.
\end{align}
This vanishes for $|\lambda|=1$ and gives
\begin{align}
V = 
\frac{\mu^2\left[4 |y|^2+(2-4|y|-6|y|^2)\left|z\right|^2+|y|^2\left|z\right|^4\right]}{(1+\left|y)\right|^2\left(2-\left|z\right|^2\right)^2} \,.
\end{align}
Expanding the potential around $z=0$ we get
\begin{align}
V=\frac{\mu^2|y|^2}{(1+|y|)^2}+\mu^2\frac{(1-|y|)^2-2|y|^2}{2(1+|y|)^2}\left|z\right|^2 + \dots\,.
\end{align}
In case we take into account the dilaton, the minimum remains at $\cos\theta=\text{sgn}(\lambda)$ and the elimination of the pole at $\left|y\right|=1$ requires again $|\lambda|=1$. Using this, the potential becomes
\begin{align}
V &\!\!=\!\! \left.e^{K} \left(\left|D_i W\right|^2\!-\!2\left|W\right|^2\right)\right|_{\cos\theta=\text{sgn}(\lambda),|\lambda|=1}\!\!=\!
\mu^2\frac{4|y|^2+(2-4|y|-2|y|^2)\left|z\right|^2+ |y|^2 \left|z\right|^4}{(S+{\bar S})\left(1+|y|\right)^2\left(2-\left|z\right|^2\right)^2}\nonumber\\
&=\frac{\hat\mu^2}{(1+|y|)^2}\left(|y|^2+\frac{1}{2}\frac{(1-|y|)^2\left|z\right|^2}{\left(1-\frac{\left|z\right|^2}{2}\right)^2}\right) = 
\frac{\hat\mu^2 |y|^2}{(1+|y|)^2} + \frac{\hat\mu^2 (1-|y|)^2}{(1+|y|)^2}\left|z\right|^2 + \dots\,,
\end{align}
where we defined $\hat\mu^2\equiv\mu^2/(S+{\bar S})$.



\end{appendices}

\bibliographystyle{utphys}
\providecommand{\href}[2]{#2}\begingroup\raggedright\endgroup

\end{document}